*Chapter 7*
# *In vivo* wireless channel modeling

*A. Fatih Demir\*, Z. Esat Ankarali\*, Yang Liu\*,*
*Qammer H. Abbasi†, Khalid Qaraqe‡, Erchin Serpedin†,*
*Huseyin Arslan\*, and Richard D. Gitlin\**

## 7.1 Introduction

Technological advances in biomedical engineering have significantly improved the quality of life and increased the life expectancy of many people. In recent years, there has been increased interest in wireless body area networks (WBANs) research with the goal of satisfying the demand for innovative biomedical technologies and improved healthcare quality [1, 2]. One component of such advanced technologies is represented by the devices such as wireless *in vivo* sensors and actuators, e.g., pacemakers, internal drug delivery devices, nerve stimulators, wireless capsule endoscopes (WCEs), etc. *In vivo* wireless medical devices and their associated technologies represent the next stage of this evolution and offer a cost efficient and scalable solution along with the integration of wearable devices. *In vivo*-WBAN devices (Figure 7.1) are capable of providing continuous health monitoring and reducing the invasiveness of surgeries. Furthermore, patient information can be collected over a larger period of time, and physicians are able to perform more reliable analysis by exploiting *big data* [3] rather than relying on the data recorded in short hospital visits [4–6].

In order to fully exploit and increase further the potential of WBANs for practical applications, it is necessary to accurately assess the propagation of electromagnetic (EM) waveforms in an *in vivo* communication environment (implant-to-implant and implant-to-external device) and obtain accurate channel models that are necessary to optimize the system parameters and build reliable, high-performance communication systems. In particular, creating and accessing such a model is necessary for achieving high data rates, target link budgets, determining optimal operating frequencies, and designing efficient antennas and transceivers including digital baseband transmitter/receiver algorithms [7, 8]. Therefore, investigation of the *in vivo* wireless communication channel is crucial to obtain better performance for *in vivo*-WBAN devices. However, research on the *in vivo* wireless communication is still in the early stages, and heretofore there have been relatively few studies compared to the on-body wireless communication channel [2, 9–11].

\*Department of Electrical Engineering, University of South Florida, Tampa, FL, USA
†Department of ECEN Texas A&M University, College Station, TX, USA
‡Department of ECEN Texas A&M University, Doha, Qatar

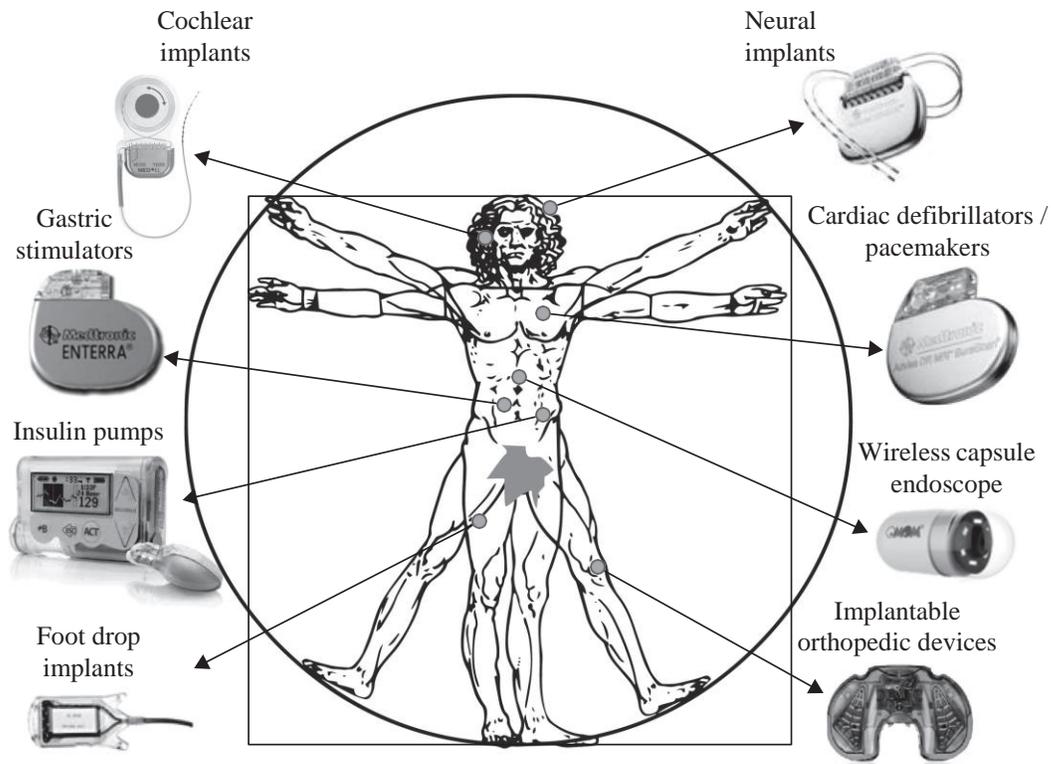

*Figure 7.1 In vivo-WBAN*

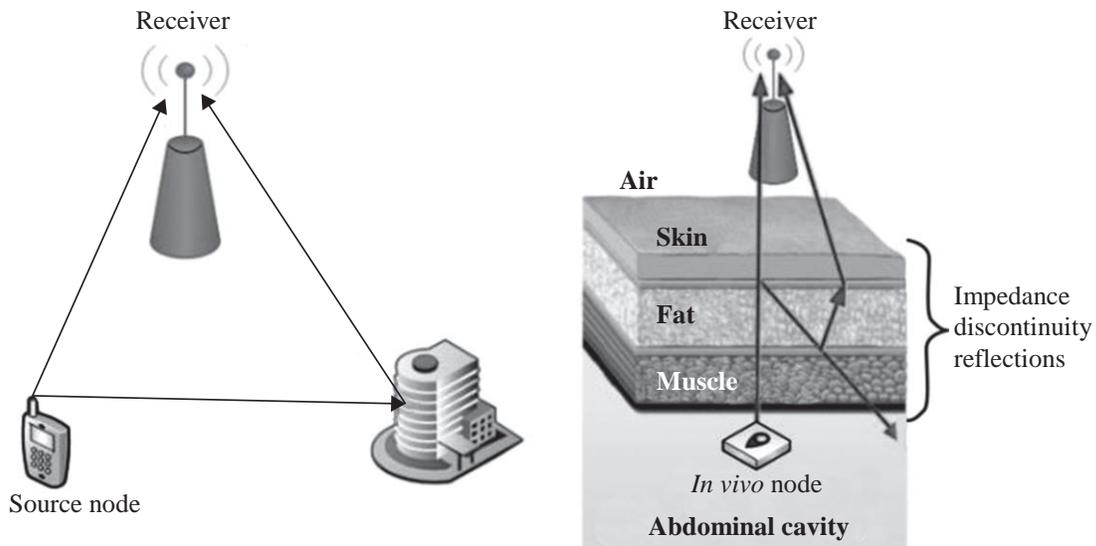

*Figure 7.2 The classical communication channel compared with the in vivo channel. ©2016 IEEE. Reprinted with permission from Reference 8*

The *in vivo* channel exhibit different characteristics than those of the more familiar wireless cellular and Wi-Fi environments since the EM wave propagates through a very lossy environment inside the body, and the predominant scatterers are present in the vicinity of the antenna (Figure 7.2). In this chapter, the state of the art of *in vivo* channel characterization is presented, and several research challenges are discussed by considering various communication methods, operational frequencies, and antenna designs. Furthermore, a numerical and experimental characterization of the *in vivo* wireless communication channel is described in detail. This chapter aims to provide a more complete picture of this fascinating communications medium and stimulate more research in this important area.

## 7.2 EM modeling of the human body

In order to investigate the *in vivo* wireless communication channel, accurate body models, and knowledge of the EM properties of the tissues are crucial [2]. Human autopsy materials and animal tissues have been measured over the frequency range 10 Hz to 20 GHz [12] and the frequency-dependent dielectric properties of the tissues are modeled using the four-pole Cole–Cole equation, which is expressed as:

$$\varepsilon(\omega) = \varepsilon_\infty + \sum_{m=1}^{4} \frac{\Delta\varepsilon_m}{1+(j\omega\tau_m)^{(1-a_m)}} + \frac{\sigma}{j\omega\varepsilon_0} \quad (7.1)$$

where $\varepsilon_\infty$ stands for the body material permittivity at terahertz frequency, $\varepsilon_0$ denotes the free-space permittivity, $\sigma$ represents the ionic conductivity and $\varepsilon_m$, $\tau_m$, $a_m$ are the body material parameters for each anatomical region. The parameters for anatomical regions are provided in Reference 13, and the EM properties such as conductivity, relative permittivity, loss tangent, and penetration depth can be derived using these parameters in (7.1).

Various physical and numerical phantoms have been designed in order to simulate the dielectric properties of the tissues for experimental and numerical investigation [14]. These can be classified as homogeneous, multilayered, and heterogeneous phantom models. Although heterogeneous models provide a more realistic approximation to the human body, design of physical heterogeneous phantoms is quite difficult and performing numerical experiments on these models is very complex and resource intensive. On the other hand, homogeneous or multilayer models cannot differentiate EM wave radiation characteristics for different anatomical regions. Figure 7.3 shows examples of heterogeneous physical and numerical phantoms.

Analytical methods are generally viewed as infeasible and require extreme simplifications. Therefore, numerical methods are used for characterizing the *in vivo* wireless communication channel. Numerical methods provide less complex and appropriate approximations to Maxwell's equations via various techniques, such as uniform theory of diffraction (UTD), method of moments (MoM), finite element method (FEM), and finite-difference time-domain method (FDTD). Each method has its own pros and cons and should be selected based on the simulation model and size, operational frequency, available computational resources, and interested characteristics, such as power delay profile (PDP), specific absorption rate (SAR), etc. A detailed comparison of these methods is available in References 2 and 15.

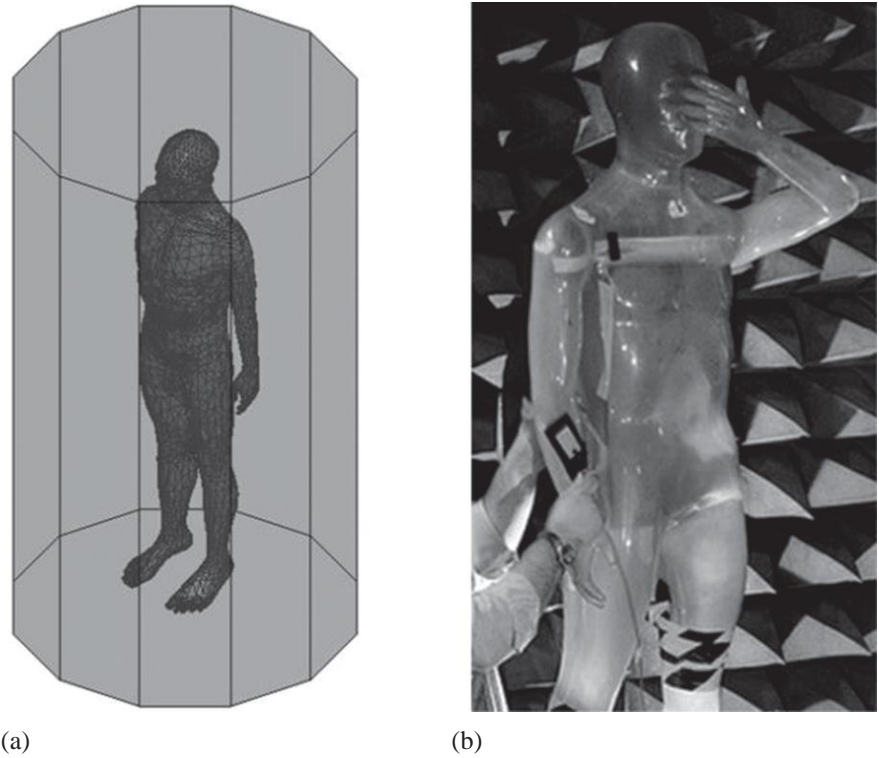

(a) (b)

*Figure 7.3 Heterogeneous human body models: (a) HFSS® model [19] and (b) physical phantom [14]. ©2016 IEEE. Reprinted with permission from Reference 55*

It may be preferable that numerical experiments should be confirmed with real measurements. However, performing experiments on a living human is carefully regulated. Therefore, anesthetized animals [16, 17] or physical phantoms, allowing repeatability of measurement results [14, 18] are often used for experimental investigation. In addition, the first such study conducted on a human cadaver was reported in Reference 20.

## 7.3 EM wave propagation through human tissues

Propagation in a lossy medium, such as human tissues, results in a high absorption of EM energy [21]. The absorption effect varies with the frequency-dependent electrical characteristics of the tissues, which mostly consist of water and ionic content [22]. The SAR provides a metric for the amount of absorbed power in the tissue and is expressed as follows [23]:

$$\text{SAR} = \frac{\sigma |E|^2}{\rho} \tag{7.2}$$

where $\sigma$, $E$, and $\rho$ represent the conductivity of the material, the RMS magnitude of the electric field, and the mass density of the material, respectively. The Federal

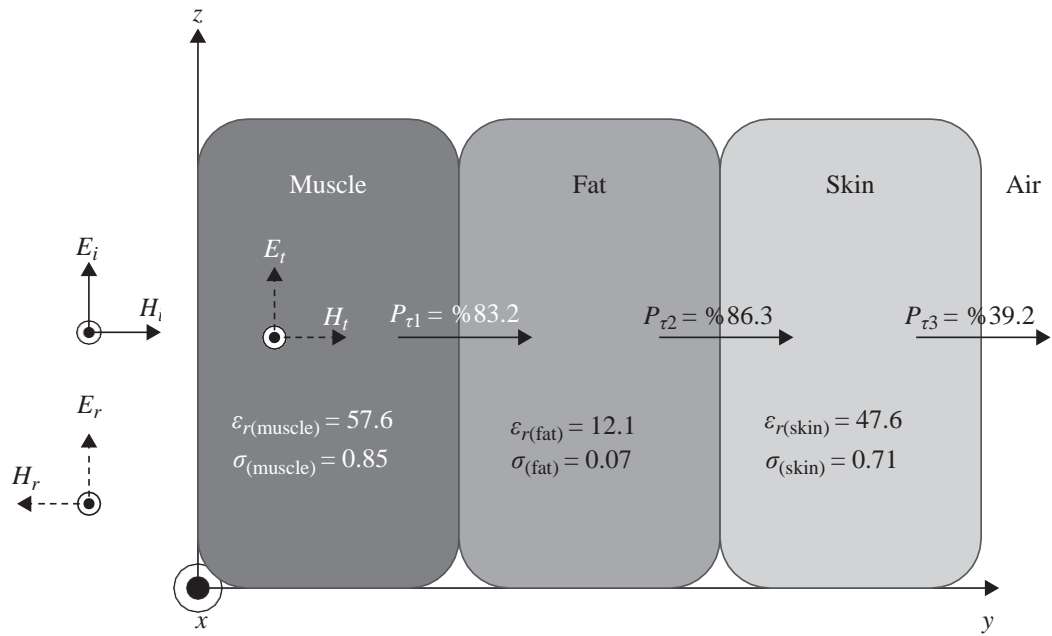

*Figure 7.4 Multi-layer human tissue model at 403 MHz ($\varepsilon_r$: permittivity, $\sigma$: conductivity, $P_\tau$: power transmission factor). ©2016 IEEE. Reprinted with permission from Reference 55*

Communications Commission (FCC) recommends the SAR to be less than 1.6 W/kg taken over a volume having 1 g of tissue [24].

When a plane EM wave propagates through the interface of two media having different electrical properties, its energy is partially reflected and the remaining portion is transmitted through the boundary of this media. Superposition of the incident and the reflected wave can cause a standing wave effect that may increase the SAR values [22]. A multilayer tissue model at 403 MHz, where each layer extends to infinity (much larger than the wavelength of EM waves) and the dielectric values are calculated using Reference 25, is illustrated in Figure 7.4. If there is a high contrast in the dielectric values of media/tissues, wave reflection at the boundary increases and transmitted power decreases.

In addition to the absorption and reflection losses, EM waves also suffer from expansion of the wave fronts (which assume an ever-increasing sphere shape from an isotropic source in free space), diffraction and scattering (which depend on the wavelength of EM wave). Section 7.6 discusses *in vivo* propagation models by considering these effects in detail.

## 7.4 Frequency of operation

Since EM waves propagate through the frequency-dependent materials inside the body, the operating frequency has an important effect on the communication channel. Accordingly, the allocated and recommended frequencies are summarized including their effects for the *in vivo* wireless communications channel in this section. The IEEE 802.15.6 standard [1] was released in 2012 to regulate short-range wireless communications inside or in the vicinity of the human body, and are classified as narrow-band (NB) communications, ultra-wide band communications (UWB), and human-body communications (HBC) [26, 27]. The frequency bands and channel bandwidths (BW) allocated for these communication methods are summarized in Table 7.1. An IEEE 802.15.6 compliant *in vivo*-WBAN device must operate in at least one of these frequency bands.

NB communications operate at lower frequencies compared to UWB communications and hence suffers less from absorption. This can be appreciated by considering (7.1) and (7.2) that describe the absorption as a function of frequency. The medical device radio communications service (MedRadio uses discrete bands within the 401–457 MHz spectrum including the international medical implant communication service (MICS) band) and medical body area network (MBAN, 2360–2400 MHz) are allocated by the FCC for medical devices usage. Therefore, co-user interference problems are less severe in these frequency bands. However, NB communications are only allocated small bandwidths (1 MHz at most) in the standard as shown in Table 7.1. The IEEE 802.15.6 standard does not define a maximum transmit power and the local regulatory bodies limit it. The maximum power is restricted to 25 W EIRP (equivalent radiated isotropic power) by FCC, whereas it is set to 25 W ERP (equivalent radiated power) by ETSI (European Telecommunication Standards Institute) for the 402–405 MHz band.

UWB communications is a promising technology to deploy inside the body due to its inherent features including high data rate capability, low power, improved penetration (propagation) abilities through tissues, and low probability of intercept. The large bandwidths for UWB (499 MHz) enable high data rate communications and applications. Also, UWB signals are inherently robust against detection and smart jamming attacks because of their extremely low maximum effective isotropic radiated power (EIRP) spectral density, which is −41.3 dBm/MHz [28, 29]. On the other hand, UWB communications inside the body suffers from pulse distortion caused by frequency-dependent tissue absorption and compact antenna design. Recently, the terahertz frequency band has also been a subject of interest for *in vivo* propagation, and it is regarded as one of the most promising bands for the EM paradigm of nano-communications [30].

Table 7.1 Frequency bands and bandwidths for the three different propagation methods in IEEE 802.15.6. ©2016 IEEE. Reprinted with permission from Reference 55

| **Propagation method** | **IEEE 802.15.6 operating freq. bands** | | **Selected references** |
|---|---|---|---|
| | **Frequency band** | **BW** | |
| Narrow band communications | 402–405 MHz<br>420–450 MHz | 300 kHz<br>300 kHz | [7, 14, 22, 31, 32, 36, 47] |
| | 863–870 MHz<br>902–928 MHz<br>950–956 MHz | 400 kHz<br>500 kHz<br>400 kHz | [7, 14, 31, 36, 45, 47] |
| | 2360–2400 MHz<br>2400–2438.5 MHz | 1 MHz<br>1 MHz | [7, 14, 36, 50] |
| UWB communications | 3.2–4.7 GHz<br>6.2–10.3 GHz | 499 MHz<br>499 MHz | [17, 28, 36, 50] |
| Human-body communications | 16 MHz<br>27 MHz | 4 MHz<br>4 MHz | [26, 27] |

## 7.5 *In vivo* antenna design considerations

Unlike free space communications, *in vivo* antennas are often considered to be an integral part of the channel, and they generally require different specifications than *ex vivo* antennas [2, 31–33]. In this section, we will describe their salient differences as compared to the *ex vivo* antennas.

*In vivo* antennas are subject to strict size constraints and in addition need to be biocompatible. Although copper antennas have better performance, only specific types of materials such as titanium or platinum should be used for *in vivo* communications due to their noncorrosive chemistry [6]. The standard definition of the gain is not valid for *in vivo* antennas since it includes body effects [34, 35]. As noted above, the gain of the *in vivo* antennas cannot be separated from the body effects as the antennas are considered to be an integral part of the channel. Hence, the *in vivo* antennas should be designed and placed carefully in order not to harm the biological tissues and to provide power efficiency. When the antennas are placed inside the human body, their electrical dimensions and gains decrease due to the high dielectric permittivity and high conductivity of the tissues, respectively [36]. For instance, fat has a lower conductivity than skin and muscle. Therefore, *in vivo* antennas are usually placed in a fat (usually subcutaneous fat –SAT–) layer to increase the antenna gain. This placement also provides less absorption losses due to shorter propagation path. However, the antenna size becomes larger in this case. In order to reduce high losses inside the tissues, a high permittivity, low loss coating layer can be used. As the coating thickness increases, the antenna becomes less sensitive to the surrounding material [36, 37].

Lossy materials covering the *in vivo* antenna change the electrical current distribution in the antenna and radiation pattern [18]. It is reported in Reference 31 that directivity of *in vivo* antennas increases due to the curvature of body surface, losses, and dielectric loading from the human body. Therefore, this increased directivity should be taken into account as well in order not to harm the tissues in the vicinity of the antenna [23].

*In vivo* antennas can be classified into two main groups as electrical and magnetic antennas. Electrical antennas, e.g., dipole antennas, generate electric fields (E-field) normal to the tissues, while magnetic antennas, e.g., loop antennas produce E-fields tangential to the human tissues [38]. Normal E-field components at the medium interfaces overheat the tissues due to the boundary condition requirements [39] as illustrated in Figure 7.5. The muscle layer has a larger permittivity value than the fat layer, and hence, the E-field increases in the fat layer. Therefore, magnetic antennas allow higher transmission power for *in vivo*-WBAN devices as can be appreciated

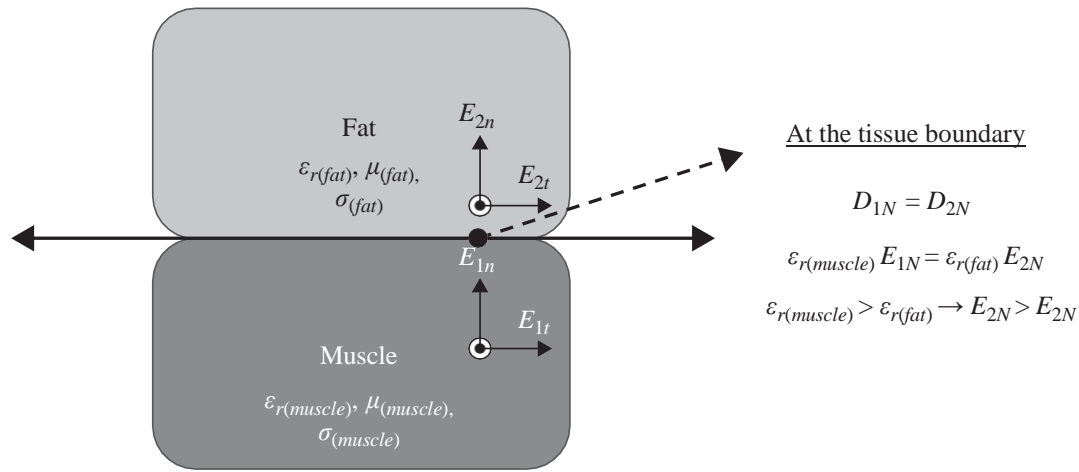

*Figure 7.5  EM propagation through tissue interface. ©2016 IEEE. Reprinted with permission from Reference 55*

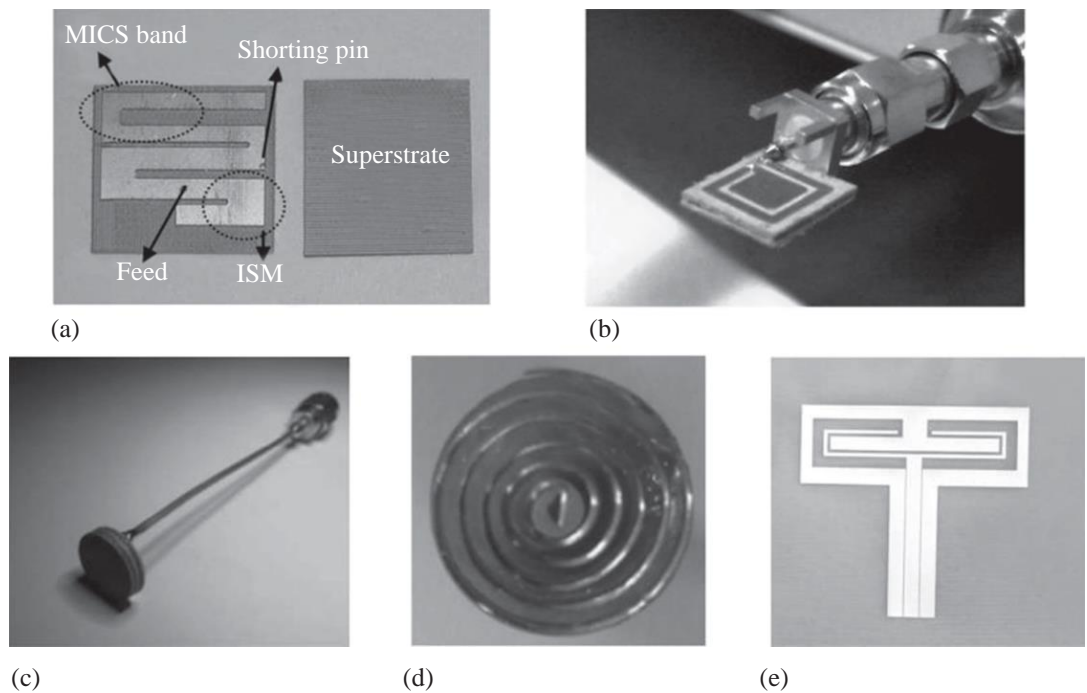

*Figure 7.6  Selected in vivo antenna samples: (a) A dual-band implantable antenna [41], (b) a miniaturized implantable broadband stacked planar inverted-F antenna (PIFA) [42], (c) a miniature scalp-implantable [43], (d) a wideband spiral antenna for WCE [16], and (e) an implantable folded slot dipole antenna [44]. ©2016 IEEE. Reprinted with permission from Reference 55*

by (7.2). In practice, magnetic loop antennas require large sizes, which is a challenge to fit inside the body. Accordingly, smaller size spiral antennas having a similar current distribution as loop antennas can be used for *in vivo* devices [40]. Several selected sample antennas designed for *in vivo* communications are shown in Figure 7.6.

## 7.6 *In vivo* EM wave propagation models

The important factors for *in vivo* wireless communication channel characterization, such as EM modeling of the human body, propagation through the tissues, and selection of the operational frequency, have been discussed in detail in the preceding sections. Further, the main differences between *in vivo* and *ex vivo* antenna designs were discussed – principally that the antenna must be considered as an integral part of the *in vivo* channel. In this section, the focus is on EM wave propagation inside the human body considering the anatomical features of organs and tissues. Then, the analytical and statistical path loss models will be presented. Since the EM wave propagates through a very lossy environment inside the body and predominant scatterers are present in the near-field region of the antenna, the *in vivo* channel exhibits different characteristics than those of the more familiar wireless cellular and Wi-Fi environments.

EM wave propagation inside the body is subject-specific and strongly related to the location of the antenna as demonstrated in References 7, 18, 20, 31, and 45. Therefore, channel characterization is generally investigated for a specific part of the human body. Figure 7.7 shows several investigated anatomical regions for various

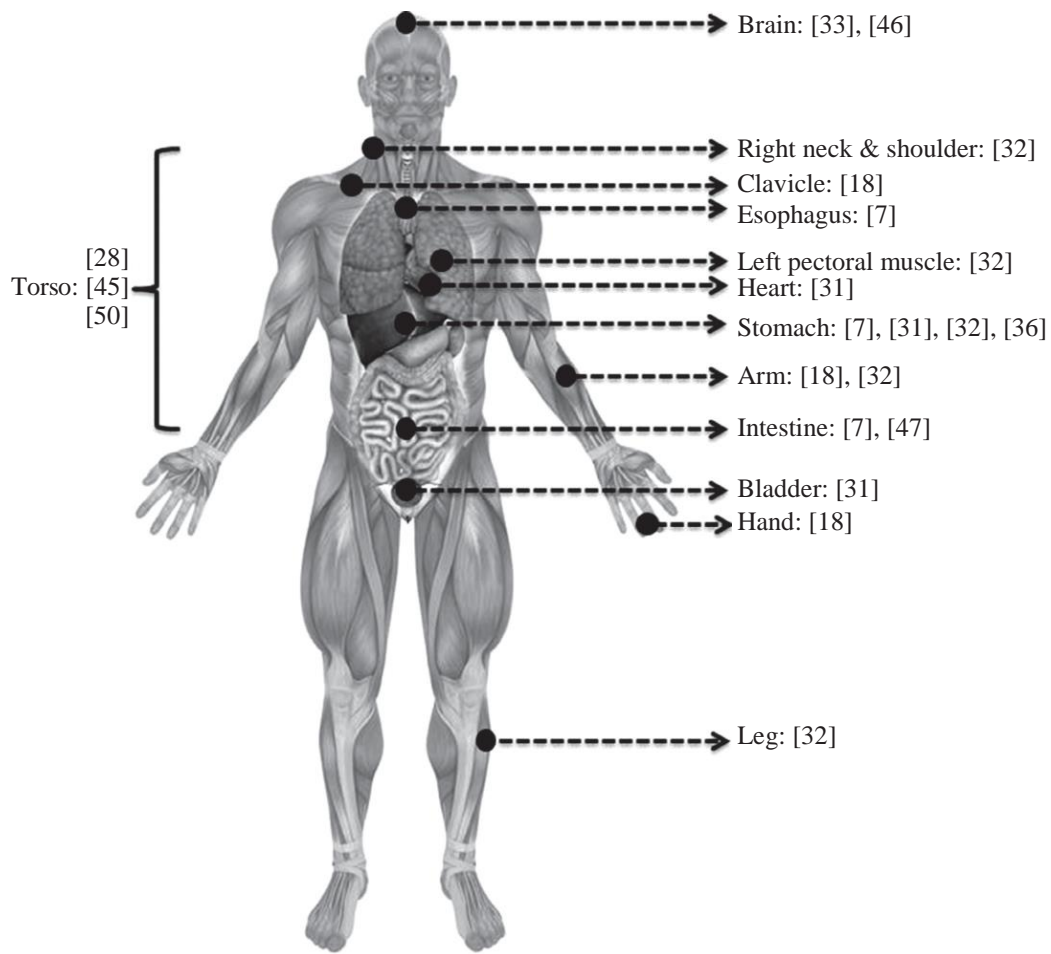

*Figure 7.7  Investigated anatomical human body regions. ©2016 IEEE. Reprinted with permission from Reference 55*

*in vivo*-WBAN applications. For example, the heart area has been studied for implantable cardioverter defibrillators and pacemakers, while the gastrointestinal (GI) tract including esophagus, stomach, and intestine has been investigated for WCE applications. The bladder region is studied for wirelessly controlled valves in the urinary tract, and the brain is investigated for neural implants [33, 46]. Also, clavicle, arm, and hands are specifically studied as they are affected less by the *in vivo* medium.

When the *in vivo* antenna is placed in an anatomically complex region, path loss, a measure of average signal power attenuation, increases [7]. This is the case with the intestine which presents a complex structure with repetitive, curvy-shaped, dissimilar tissue layers, while the stomach has a smoother structure. As a result, the path loss is greater in the intestine than in the stomach even at equal *in vivo* antenna depths [7].

Various analytical and statistical path loss formulas have been proposed for the *in vivo* channel in the literature as listed in Table 7.2. These formulas have been derived considering different shadowing phenomena for the *in vivo* medium. The initial three models are functions of the Friis transmission equation [51], return loss, and absorption in the tissues. These models are valid, when either the far field conditions are fulfilled or few scattering objects exist between the transmitter and receiver antennas. In the first model, the free space path loss (FSPL) is expressed by the Friis transmission

Table 7.2   *A review of selected studied path loss models for various scenarios. ©2016 IEEE. Reprinted with permission from Reference 55*

| Model | Formulation |
|---|---|
| FSPL [31] | $P_r = P_t G_t G_r \left(\dfrac{\lambda}{4\pi R}\right)^2$ |
| FSPL with RL [31], [36] | $P_r = P_t G_t (1 - |S_{11}|^2) G_r (1 - |S_{22}|^2) \left(\dfrac{\lambda}{4\pi R}\right)^2$ |
| FSPL with RL and absorption [40] | $P_r = P_t G_t (1 - |S_{11}|^2) G_r (1 - |S_{22}|^2) \left(\dfrac{\lambda}{4\pi R}\right)^2 (e^{-aR})^2$ |
| PMBA for near and far fields [48] | $P_{rn} = \dfrac{16\delta(P_t - P_{NF})}{\pi L^2} A_e,\ P_{rf} = \left(\dfrac{(P_t - P_{NF} - P_{FF})\lambda^2}{4\pi R^2}\right) G_t G_r$ |
| Statistical model – A [45], [50] | $PL(d) = PL_0 + n(d/d_0) + S(d_0 \le d)$ |
| Statistical model – B [14], [31], [32] | $PL(d) = PL(d_0) + 10n \log_{10}(d/d_0) + S(d_0 \le d)$ |

$P_r/P_t$, stands for the received/transmitted power; $G_r/G_t$ denotes the gain of the receiver/transmitter antenna; $\lambda$ represents the free space wavelength; $R$ is the distance between transmitter and receiver antennas, $|S_{11}|$ and $|S_{22}|$ stand for the reflection coefficient of receiver/transmitter antennas; $a$ is the attenuation constant, $P_{NF}/P_{FF}$ is the loss in the near/far fields; $\delta$ is $A_e/A$ where $A_e$ is the effective aperture and $A$ is the physical aperture of the antenna; $L$ is the largest dimension of the antenna; $d$ is the depth distance from the body surface, $d_0$ is the reference depth distance, $n$ is the path loss exponent; $PL_0$ is the intersection term in dB; $S$ denotes the random shadowing term. Abbreviations: FSPL represents the free space path loss in the far field, RL is the return loss, and PMBA denotes the propagation loss model.

equation. FSPL mainly depends on the gain of antennas, distance, and operating frequency. Its dependency on distance is a result of expansion of the wave fronts as explained in Section 7.3. Additionally, FSPL is frequency dependent due to the relationship between the effective area of the receiver antenna and wavelength. The two equations of the FSPL model in Table 7.2 are derived including the antenna return loss and absorption in the tissues. Another analytical model, PMBA [48], calculates the SAR over the entire human body for the far and near fields and gives the received power using the calculated absorption. Although these analytical expressions provide intuition about each component of the propagation models, they are not practical for link budget design as is the case with the wireless cellular communication environment.

The channel modeling subgroup (Task Group 15.6), which worked on developing the IEEE 802.15.6 standard, submitted their final report on body area network (BAN) channel models in November 2010. In this report, it is determined that the Friis transmission equation can be used for *in vivo* scenarios by adding a random variation term, and the path loss is modeled statistically with a log-normal distributed random shadowing $S$ and path loss exponent $n$ [29, 49]. The path loss exponent ($n$) heavily depends on environment and is obtained by performing extensive simulations and measurements. In addition, the shadowing term ($S$) depends on the different body materials (e.g., bone, muscle, fat, etc.) and the antenna gain in different directions [32]. The research efforts on assessing the statistical properties of the *in vivo* propagation channel are not finalized. There are still many open research efforts dedicated to building analytical models for different body parts and operational frequencies [14, 20, 31, 32, 45, 50].

## 7.7 *In vivo* channel characterization

The numerical *in vivo* channel characterization was performed in [45] using ANSYS HFSS® 15.0, which is a full-wave EM field simulator based on the FEM. ANSYS also provides a detailed human body model of adult male. The numerical investigation was validated by conducting experiments on a human cadaver in a laboratory environment [20]. Istanbul Medipol University provided the ethical approval for the study and medical assistance for this study.

### 7.7.1 Simulation setup

The simulations [45] have been designed based on an implant-to-external device (in-body to on-body) communications scenario. The human male torso area was divided into four sub regions considering the major internal organs: heart, stomach, kidneys, and intestine as shown in Figure 7.8. The measurements were performed in each sub region by rotating both the receiver (*ex vivo*) and transmitter antennas (*in vivo*) around the body on a plane at 22.5° angle increments as shown in Figure 7.9. For each location of the *ex vivo* antenna (5 cm away from the body surface), the *in vivo* antenna was placed at 10 different depths (10–100 mm). Moreover, the antennas were placed in the same direction in order to prevent polarization losses.

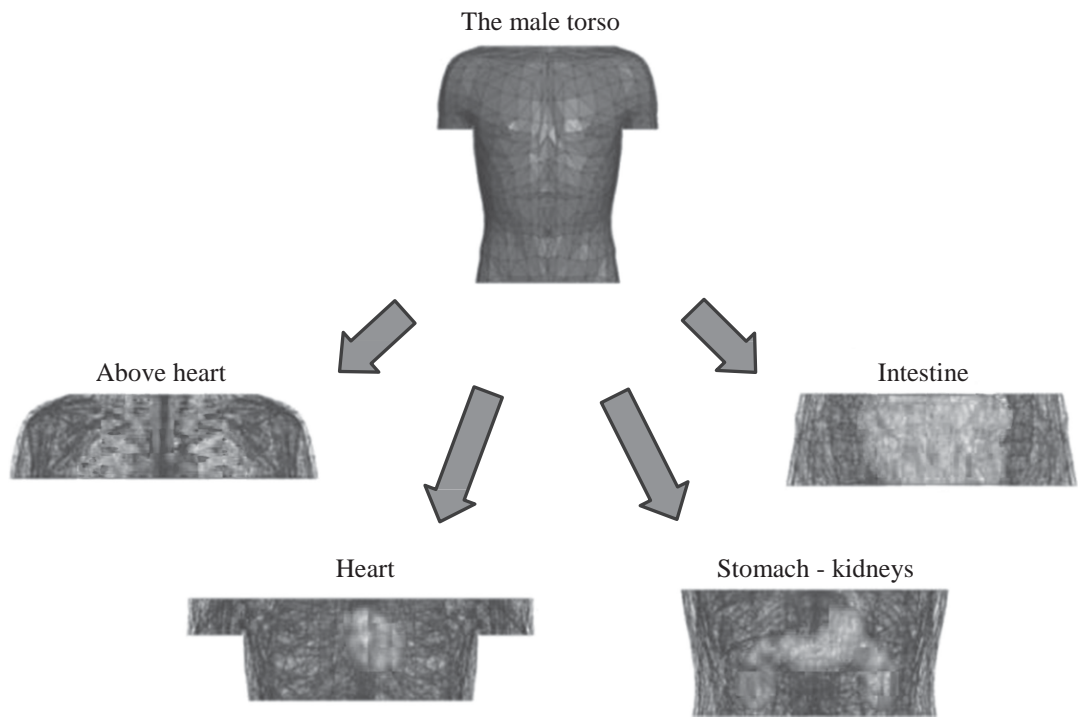

*Figure 7.8   Investigated anatomical human body regions. ©2014 IEEE. Reprinted with permission from Reference 45*

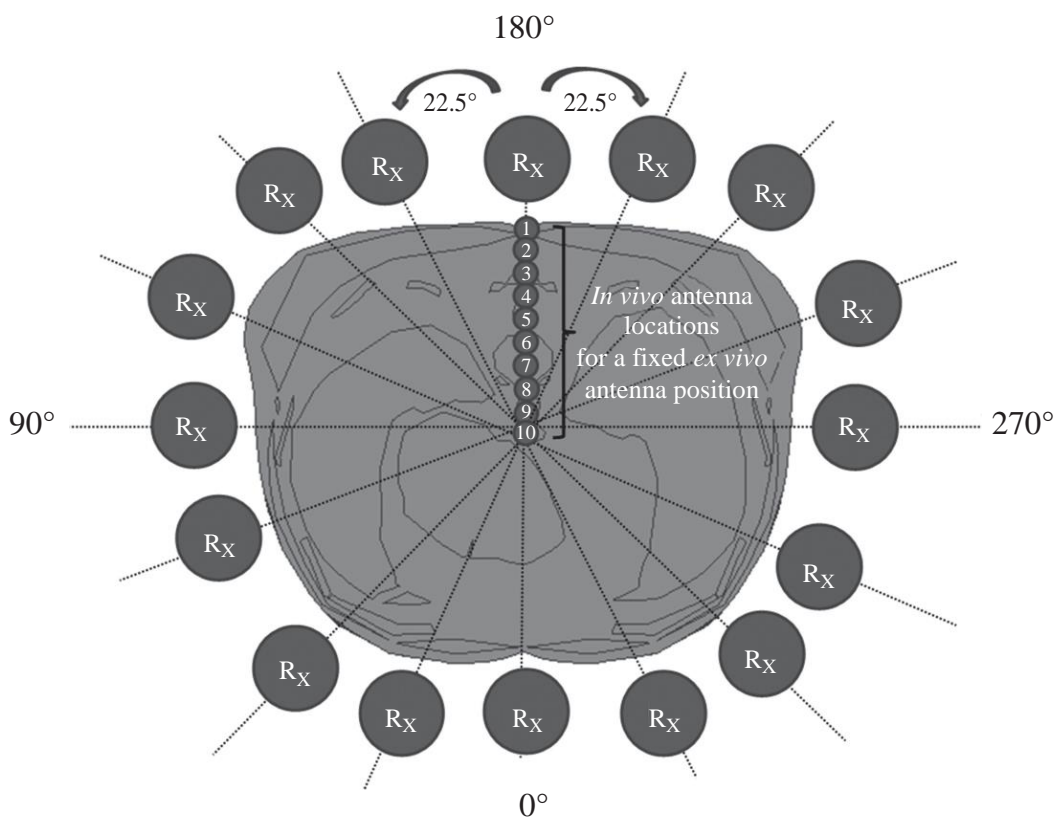

*Figure 7.9   In vivo and ex vivo antenna locations in the simulation.*

*16 (angles) × 10 (depth) × 4 (regions) = 640 simulations were performed for path loss in total. ©2014 IEEE. Reprinted with permission from Reference 45*

Omni-directional dipole antennas at 915 MHz were deployed in simulations for simplicity. The dipole antenna size is proportional to the wavelength, which changes with respect to both frequency and permittivity. Although the frequency of operation was fixed in this study, the permittivity of the environment was variable. Therefore, the antennas were optimized inside the body with respect to the average torso permittivity in order to obtain maximum power delivery. In addition, a few antenna locations with high return loss (i.e., $> -7$ dB) discarded from the data.

### 7.7.2 Experimental Setup

In order to validate our simulation results in [45], we conducted experiments on a human cadaver with a similar setup [20]. The human male torso area is investigated at 915 MHz by measuring the channel response through a vector network analyzer (VNA), while using two antennas, one (*in vivo*) [52], and other a dipole antenna (*ex vivo*) as illustrated in Figure 7.10. The *in vivo* antenna was placed at six different locations (Figure 7.11) inside the body around heart, stomach, and intestine by a physician. The antennas were located in the same orientation, and all return loss values were less than $-7$ dB in the experiment dataset.

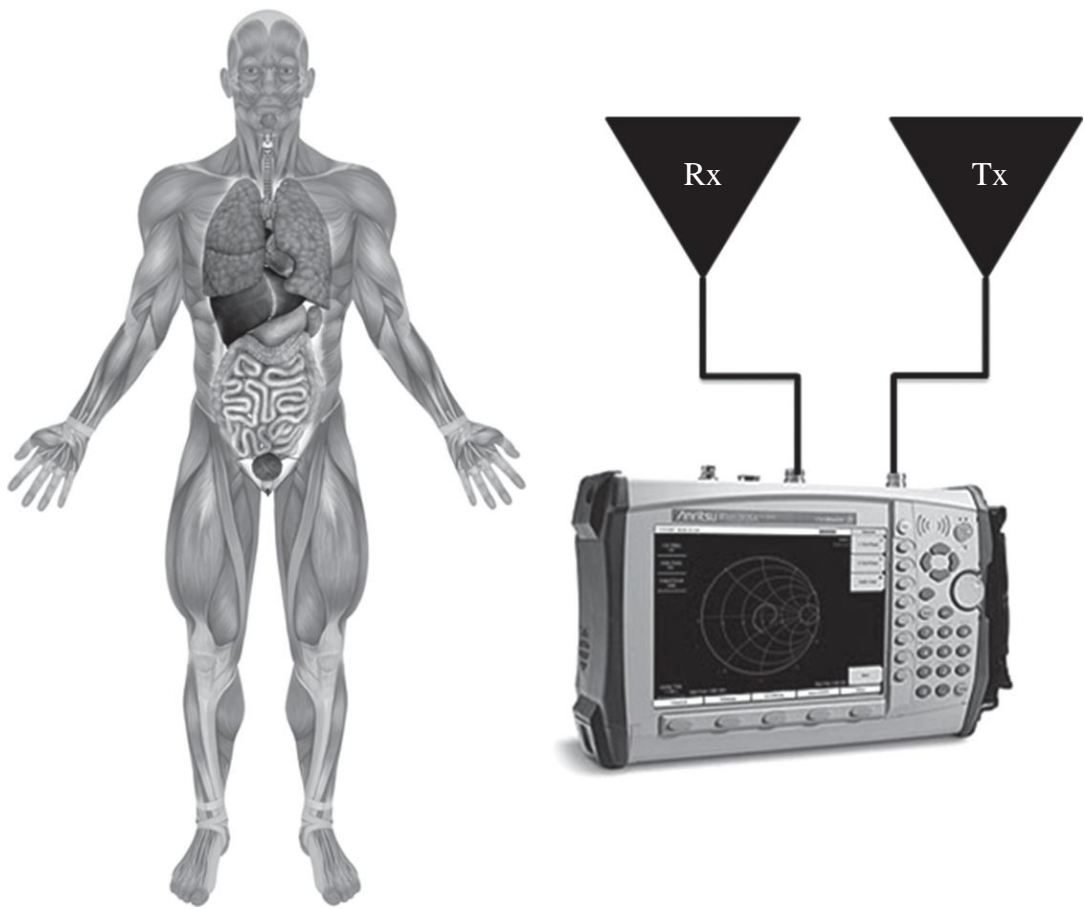

*Figure 7.10* Experiment setup for in vivo channel. ©2015 IEEE. Reprinted with permission from Reference 20

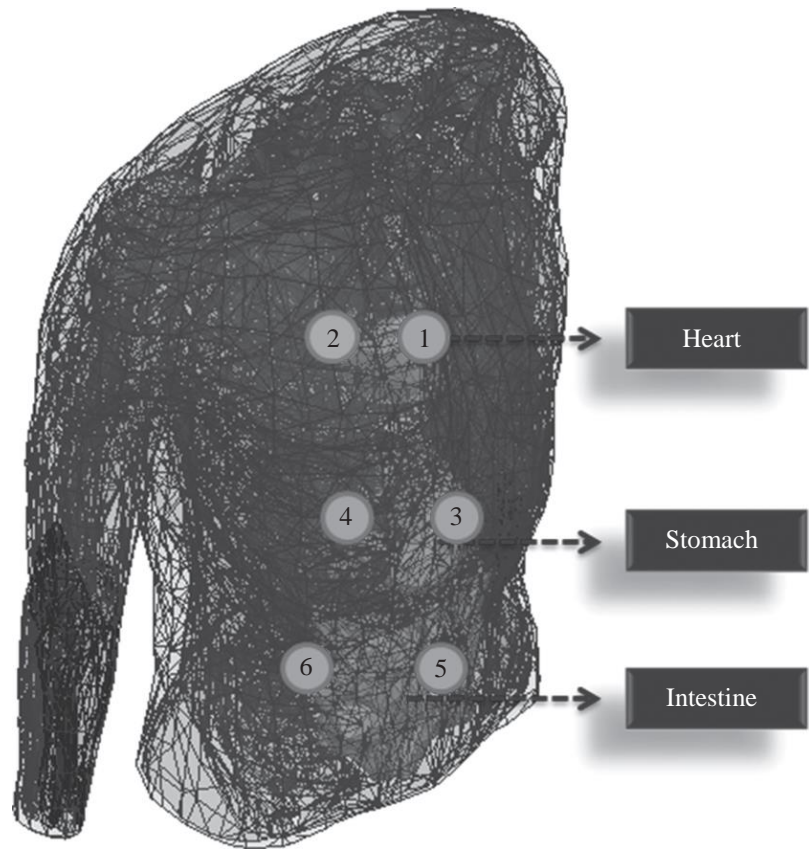

*Figure 7.11  Measurement locations on human cadaver. ©2015 IEEE. Reprinted with permission from Reference 20*

### 7.7.3  Results

#### 7.7.3.1  Location-dependent characteristics

The location-dependent characteristics of the *in vivo* path loss have investigated at 915 MHz. The EM signal propagates through different organs and tissues for various antenna locations that the path loss varies significantly even for equal *in vivo* depths. Figure 7.12 presents the mean path loss for each angular position (see Figure 7.9). It is observed that 0° has the highest path loss, whereas symmetric locations, 112.5° and 247.5°, have the lowest attenuation. In addition, the number of scattering objects (random variables) increases as the *in vivo* antenna is placed deeper and the variance of path loss increases as well due to summation of random variables.

Figure 7.13 shows the scatter plot of path loss versus depth, and the *in vivo* path loss is modeled statistically as a function of depth by the following equation in dB:

$$PL(d) = PL_0 + m(d/d_0) + S(d_0 \leq d) \tag{7.3}$$

where $d$ is the depth distance from the body surface in millimeters, $d_0$ stands for the reference depth distance (i.e., 10 mm), $PL_0$ represents the intersection term in dB, $m$ denotes the decay rate of received power, and $S$ is the shadowing term in dB, which is a normally distributed random variable with zero mean and variance $\sigma$. The parameters for the statistical *in vivo* path loss model are provided in Tables 7.3 and 7.4. There exists a 30% difference between the received power decay rates ($m$) of heart and stomach areas. In addition, the path loss at heart and intestine areas exhibits more deviation around the mean than other two regions. It could be concluded that the path loss increases significantly, when the *in vivo* antenna is placed in an anatomically complex region as also reported in Reference 7.

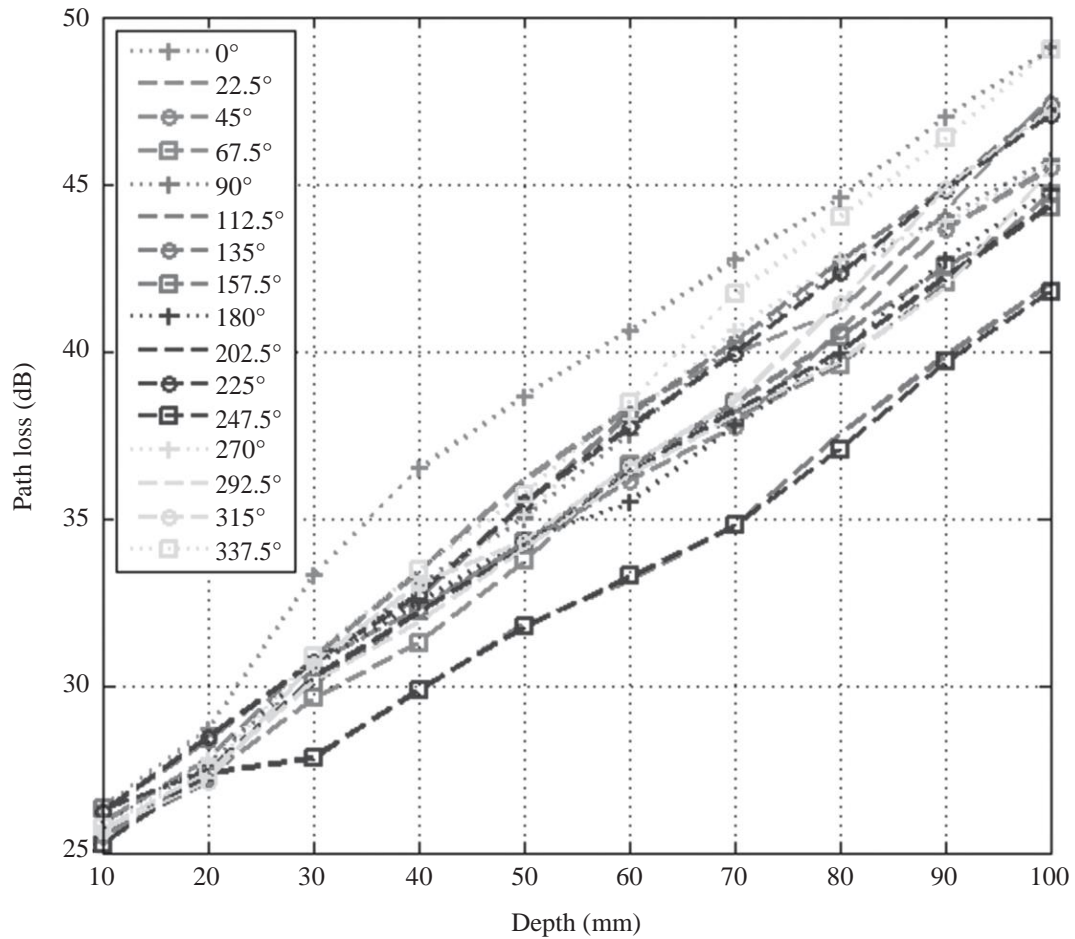

*Figure 7.12   Path loss versus in vivo depth at 915 MHz. ©2014 IEEE. Reprinted with permission from Reference 45*

The numerical studies were validated with experiments on human cadaver at 915 MHz. The *in vivo* antennas were placed at six different locations as shown in Figure 7.11, and the *ex vivo* antenna was placed 2 cm away from the body surface. Table 7.5 presents the path loss values for the selected *in vivo* locations, and a comparison of experimental results with numerical studies is provided in Figure 7.14. The discrepancies should have occurred due to additional losses which are not considered in simulations.

The angular-dependent characteristics of the *in vivo* channel were investigated by performing further simulations at 0.4 GHz, 1.4 GHz, and 2.4 GHz. The *in vivo* antenna was fixed inside the abdomen (78 mm in depth from body surface), and the *ex vivo* antenna was rotated on the body surface with the azimuth angle of 0°–355° with 5° increment. The results are presented in Figure 7.15 and Table 7.6. It could be observed that the angular dependency (i.e., the variation of the path loss versus azimuth angle) in terms of peak to average ratio is similar for different frequencies.

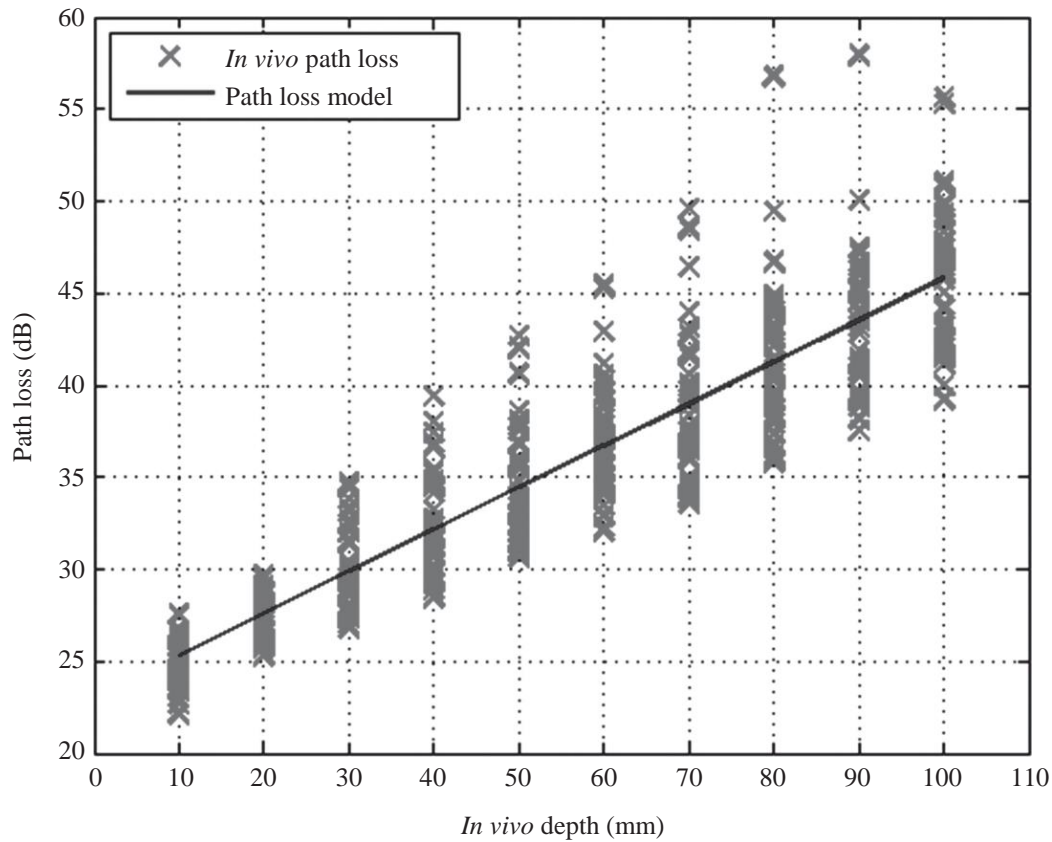

*Figure 7.13  Path loss versus in vivo depth at 915 MHz. ©2014 IEEE. Reprinted with permission from Reference 45*

*Table 7.3  Parameters for the statistical path loss model (body region). ©2014 IEEE. Reprinted with permission from Reference 45*

| Parameters ʼ Body area | $PL_0$[dB] | m | σ |
| --- | --- | --- | --- |
| Above heart | 24.75 | 2.30 | 3.73 |
| Heart | 22.70 | 1.96 | 2.38 |
| Stomach–kidneys | 22.56 | 2.55 | 1.79 |
| Intestine | 24.23 | 2.31 | 3.47 |
| Overall torso area | 23.56 | 2.28 | 3.38 |

*Table 7.4  Parameters for the statistical path loss model (body side). ©2014 IEEE. Reprinted with permission from Reference 45*

| Parameters ʼ Body area | $PL_0$[dB] | m | σ |
| --- | --- | --- | --- |
| Anterior | 23.83 | 2.46 | 3.51 |
| Posterior | 23.76 | 2.21 | 1.92 |
| Left lateral | 23.34 | 2.28 | 3.67 |
| Right lateral | 23.22 | 2.27 | 3.51 |
| Overall torso area | 23.56 | 2.28 | 3.38 |

Table 7.5 Path loss values for selected in vivo locations. ©2015 IEEE. Reprinted with permission from Reference 20

| Location | *In vivo* depth (cm) | Path loss (dB) |
| --- | --- | --- |
| 1) Above heart | 3 | 45.32 |
| 2) Below heart | 8 | 55.61 |
| 3) Above stomach | 5 | 48.19 |
| 4) Inside stomach | 9 | 50.80 |
| 5) Above intestine | 2 | 29.95 |
| 6) Below intestine | 10 | 50.47 |

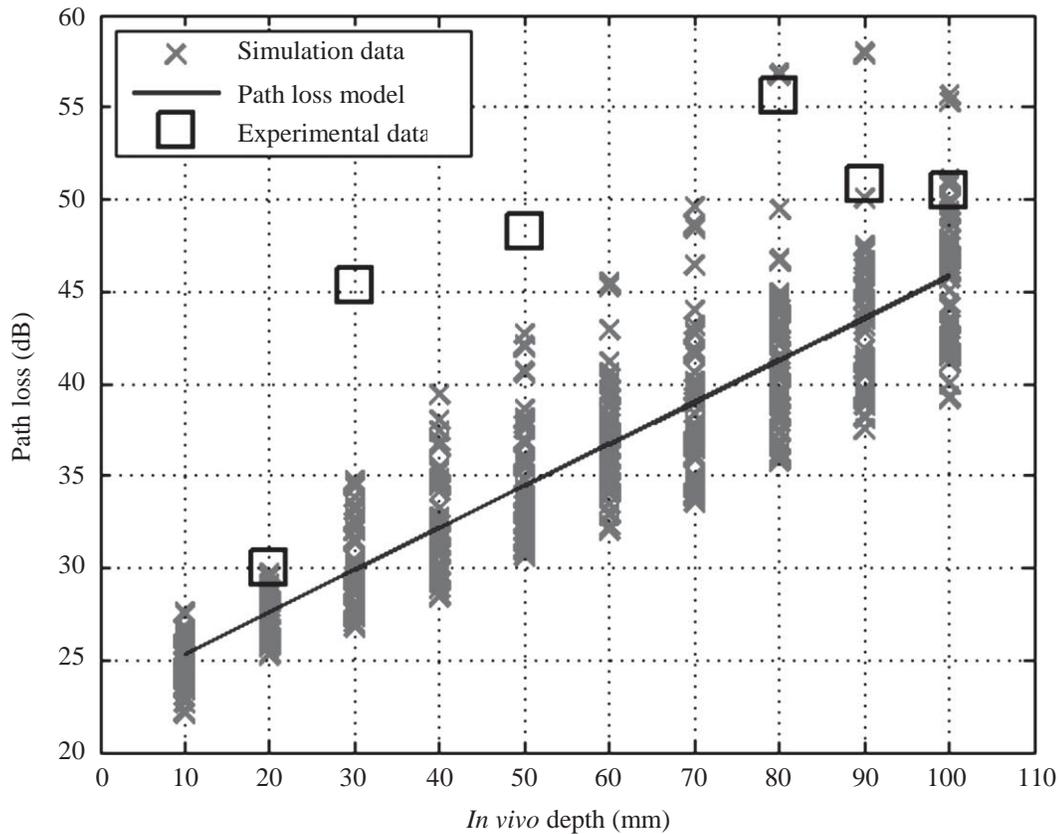

Figure 7.14 Path loss versus in vivo depth from body surface. ©2015 IEEE. Reprinted with permission from Reference 20

### 7.7.3.2 Frequency-dependent characteristics

Since the EM waves propagate through the frequency-dependent materials inside the body, the operating frequency has an important effect on the path loss model as well. The frequency-dependent characteristics of the *in vivo* channel were investigated by performing simulations from 0.4 GHz to 6 GHz at 0.1 GHz increment. The *in vivo* antenna was implanted in the abdomen (78 mm in depth from the body surface), and the *ex vivo* antenna was placed at three different locations: $d = 50$ mm (in body), $d = 78$ mm (on body), $d = 200$ mm (far external node), where $d$ denotes the distance between the transmitter and receiver. The path loss was measured for these three scenarios (implant to implant, implant to on-body, and implant to far external node), and the results are plotted in Figure 7.16 [53]. It is observed that the frequency-dependent path loss [in dB] increases linearly. Therefore, the frequency-dependent *in vivo* path loss [in ratio] increases exponentially, which is faster than that in free space.

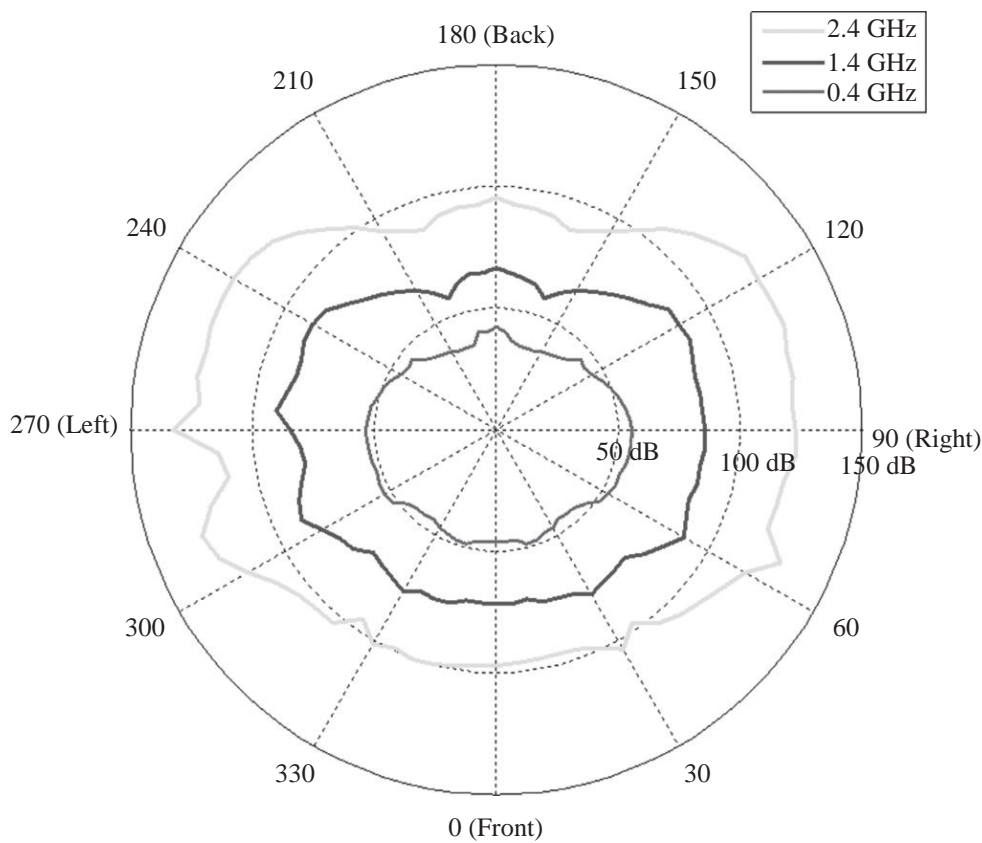

Figure 7.15  Angular-dependent path loss for on body receiver

Table 7.6  Comparison of angular-dependent path loss at different frequencies

| Frequencies (GHz) | 0.4 | 1.4 | 2.4 |
| --- | --- | --- | --- |
| Average (dB) | 46.316 | 76.74442 | 108.8819 |
| Maximum difference (dB) | 20.3373 | 33.04337 | 45.38211 |
| Peak to average ratio | 1.197665 | 1.171730 | 1.208047 |

### 7.7.3.3  Time dispersion characteristics

In addition to the path loss, the time dispersion characteristics of the *in vivo* channel were investigated for different body regions using a PDP in the simulation environment as shown in Figure 7.17. It is observed that greater dispersion is present in the sides than anterior or posterior body locations at 915 MHz. Interestingly, the torso area exhibits an exponential decaying behavior on the dB scale while linear decaying is observed for the classical indoor/outdoor channel models [28]. The maximum excess delay is not more than 10 ns which might be negligible for NB communications. However, for UWB communications, which is also a very popular signaling scheme in WBAN research, this dispersion may lead to a significant interference effect and should be carefully considered in the waveform design.

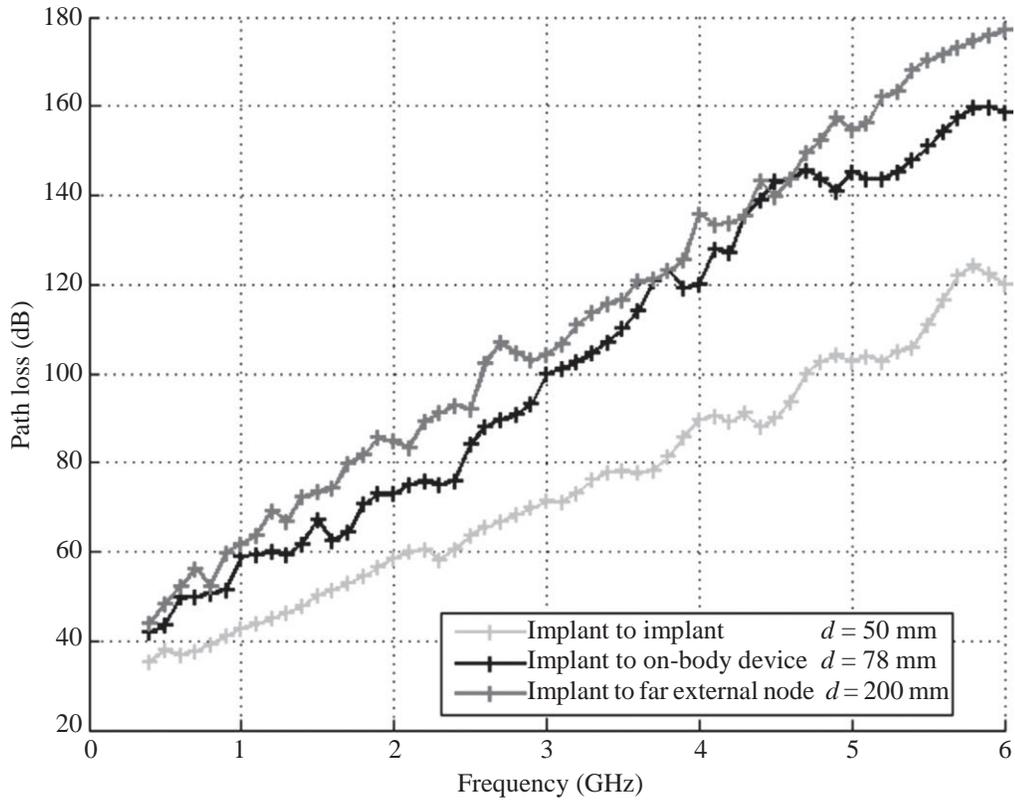

*Figure 7.16 Frequency-dependent path loss at different locations. ©2015 IEEE. Reprinted with permission from Reference 53*

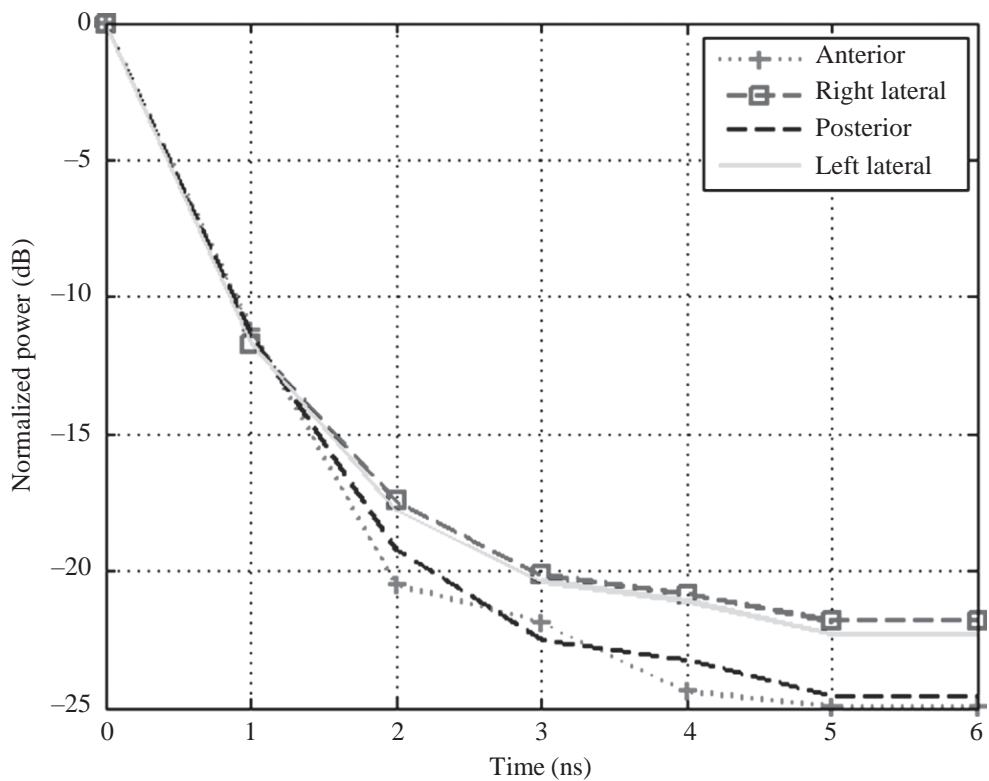

*Figure 7.17 Power delay profile for each body side. ©2014 IEEE. Reprinted with permission from Reference 45*

## 7.8 Comparison of *in vivo* and *ex vivo* channels

We summarize the differences between the *in vivo* and *ex vivo* channels in Table 7.7.

Table 7.7  Comparison of in vivo and ex vivo channels. ©2015 IEEE. Reprinted with permission from Reference 54

| **Feature** | *Ex vivo* | *In vivo* |
|---|---|---|
| Physical wave propagation | Constant speed<br>Multipath | Variable speed<br>Multipath – plus penetration in biological tissues |
| Attenuation and path loss | Lossless medium (losses are negligible)<br>Path loss is essentially uniform<br>Increases with distance | Very lossy medium<br>Location dependent<br>Increases exponentially with distance inside the body |
| Dispersion | Multipath delays → time dispersion | Multipath delays of variable speed → frequency-dependent time dispersion |
| Near-field communications | Deterministic near-field region around the antenna | Inhomogeneous medium → near-field region changes with angles and position inside body |
| Power limitations | Average and peak | Average and peak – plus SAR |
| Shadowing | Follows a *log-normal* distribution | Follows a *log-normal* distribution |
| Multipath fading | Flat/frequency selective fading | Lower speed of propagation causes longer dispersion than in free space |
| Antenna | Antenna gain is essentially location independent | "Implant location"-dependent antenna gain |
| Wavelength | In free space → the speed of light divided by operational frequency | $\lambda = \frac{c}{\sqrt{\varepsilon_r f}}$ (e.g., $\varepsilon_r = 35$ at 2.4 GHz → roughly six times shorter than the wavelength in free space) |

## 7.9 Summary

In this chapter, the state of the art of *in vivo* wireless channel characterization has been presented. Various studies described in the literature are dedicated to the *in vivo* communication channel, and they consider different parameters in studying various anatomical regions. Furthermore, the location-dependent characteristics of *in vivo* wireless communication at 915 MHz are analyzed in detail via numerical and experimental investigations. A complete model for the *in vivo* channel is not available and remains an open research problem. However, considering the expected future growth of implanted technologies and their potential use for the detection and diagnosis of various health-related issues in the human body, the channel modeling studies should be further extended to develop better and more efficient communications systems for future *in vivo* systems.

## Acknowledgements

This publication was made possible by NPRP grant # 6-415-3-111 from the Qatar National Research Fund (a member of Qatar Foundation). The statements made herein are solely the responsibility of the authors.

## References

[1] "IEEE standard for local and metropolitan area networks: Part 15.6: Wireless body area networks," IEEE submission, February 2012, IEEE Std.
[2] P. S. Hall and Y. Hao, Antennas and Propagation for Body-Centric Wireless Communications, 2nd Edition. Norwood, MA: Artech House, 2012.
[3] S. Yu, X. Lin, and J. Misic, "Networking for big data [guest editorial]," Network, IEEE, vol. 28, no. 4, pp. 4–4, July–August 2014.
[4] A. Taparugssanagorn, A. Rabbachin, M. Hamalainen, J. Saloranta, and J. Iinatti, "A review of channel modelling for wireless body area network in wireless medical communications," in Proceeding of the 11th International Symposium on Wireless Personal Multimedia Communications (WPMC), 2008.
[5] A. Kiourti, K. A. Psathas, and K. S. Nikita, "Implantable and ingestible medical devices with wireless telemetry functionalities: A review of current status and challenges," Bioelectromagnetics, vol. 15, pp. 1–15, August 2013.
[6] S. Movassaghi, M. Abolhasan, J. Lipman, D. Smith, and A. Jamalipour, "Wireless body area networks: A survey," Communication Surveys and Tutorials, IEEE, vol. 16, pp. 1–29, 2014.
[7] M. R. Basar, F. Malek, K. M. Juni, *et al.*, "The use of a human body model to determine the variation of path losses in the human body channel in wireless capsule endoscopy," Progress in Electromagnetics Research, vol. 133, pp. 495–513, 2013.

[8] T. P. Ketterl, G. E. Arrobo, A. Sahin, T. J. Tillman, H. Arslan, and R. D. Gitlin, "*In vivo* wireless communication channels," in IEEE 13th Annual Wireless and Microwave Technology Conference (WAMICON), 2012.

[9] D. Smith, D. Miniutti, T. Lamahewa, and L. Hanlen, "Propagation models for body-area networks: A survey and new outlook," Antennas and Propagation Magazine, IEEE, vol. 55, pp. 97–117, Oct. 2013.

[10] Q. H. Abbasi, A. Sani, A. Alomainy, and Y. Hao, "On-body radio channel characterisation and system-level modelling for multiband OFDM ultra wideband body-centric wireless network," IEEE Transactions on Microwave Theory and Techniques, vol. 58, no. 12, pp. 3485–3492, December 2010.

[11] Q. H. Abbasi, A. Sani, A. Alomainy, and Y. Hao, "Numerical characterisation and modelling of subject-specific ultra wideband body-centric radio channels and systems for healthcare applications," IEEE Transaction on Information and Technology in Biomedicine, vol. 16, no. 2, pp. 221–227, March 2012.

[12] S. Gabriel, R. Lau, and C. Gabriel, "The dielectric properties of biological tissues: II. Measurements in the frequency range 10 Hz to 20 GHz," Physics in Medicine and Biology, vol. 40, no. 11, p. 2251, 1996.

[13] S. Gabriel, R. W. Lau, and C. Gabriel, "The dielectric properties of biological tissues: III. Parametric models for the dielectric spectrum of tissues," Physics in Medicine and Biology, vol. 41, pp. 2271–2293, 1996.

[14] A. Alomainy and Y. Hao, "Modeling and characterization of biotelemetric radio channel from ingested implants considering organ contents," IEEE Transactions on Antennas and Propagation, vol. 57, pp. 999–1005, April 2009.

[15] A. Pellegrini, A. Brizzi, L. Zhang, *et al.*, "Antennas and propagation for body-centric wireless communications at millimeter-wave frequencies: A review [wireless corner]," Antennas and Propagation Magazine, IEEE, vol. 55, no. 4, pp. 262–287, 2013.

[16] S. H. Lee, J. Lee, Y. J. Yoon, *et al.*, "A wideband spiral antenna for ingestible capsule endoscope systems: Experimental results in a human phantom and a pig," IEEE Transactions on Biomedical Engineering, vol. 58, no. 6, pp. 1734–1741, June 2011.

[17] R. Chavez-Santiago, I. Balasingham, J. Bergsland, *et al.*, "Experimental implant communication of high data rate video using an ultra wideband radio link," in Engineering in Medicine and Biology Society (EMBC), 2013 35th Annual International Conference of the IEEE. IEEE, 2013, pp. 5175–5178.

[18] H.-Y. Lin, M. Takahashi, K. Saito, and K. Ito, "Characteristics of electric field and radiation pattern on different locations of the human body for in-body wireless communication," IEEE Transactions on Antennas and Propagation, vol. 61, pp. 5350–5354, October 2013.

[19]   http://www.ansys.com/Products.

[20] A. F. Demir, Q. H. Abbasi, Z. E. Ankarali, M. Qaraqe, E. Serpedin and H. Arslan, "Experimental Characterization of *In Vivo* Wireless Communication Channels," Vehicular Technology Conference (VTC Fall), 2015 IEEE 82nd, Boston, MA, 2015, pp. 1–2.

[21] B. Latre, B. Braem, I. Moerman, C. Blondia, and P. Demeester, "A survey on wireless body area networks," Wireless Networks, vol. 17, pp. 1–18, November 2010.
[22] K. Y. Yazdandoost, Wireless Mobile Communication and Healthcare. Springer Berlin Heidelberg, 2012, ch. A Radio Channel Model for In-body Wireless Communications, pp. 88–95.
[23] "C95.1-200S: IEEE Standard for Safety Levels With Respect to Human Exposure to Radio Frequency Electromagnetic Fields, 3 kHz to 300 GHz," 2006, IEEE Std.
[24] T. P. Ketterl, G. E. Arrobo, and R. D. Gitlin, "SAR and BER evaluation using a simulation test bench for *in vivo* communication at 2.4 GHz," in Wireless and Microwave Technology Conference (Wamicon), IEEE, 2013.
[25] W. G. Scanlon, "Analysis of tissue-coupled antennas for UHF intra-body communications," *Antennas and Propagation, (ICAP 2003)*. Twelfth International Conference on (Conf. Publ. No. 491), vol. 2, 2003, pp. 747–750.
[26] M. S. Wegmueller, A. Kuhn, J. Froehlich, *et al.*, "An attempt to model the human body as a communication channel," IEEE Transactions on Biomedical Engineering, vol. 54, no. 10, pp. 1851–1857, 2007.
[27] A. T. Barth, M. A. Hanson, H. C. Powell Jr, D. Unluer, S. G. Wilson, and J. Lach, "Body-coupled communication for body sensor networks," in Proceedings of the ICST 3rd international conference on Body area networks. ICST (Institute for Computer Sciences, Social-Informatics and Telecommunications Engineering), 2008, p. 12.
[28] R. C.-S. Khaleghi and I. Balasingham, "Ultra-wideband statistical propagation channel model for implant sensors in the human chest," IET Microwaves, Antennas & Propagation, vol. 5, p. 1805, 2011.
[29] R. Chavez-Santiago, K. Sayafian Pour, A. Khaleghi, *et al.*, "Propagation models for IEEE 802.15.6 standardization of implant communication in body area networks," in IEEE Communications Magazine, vol. 51, no. 8, pp. 80–87, August 2013.
[30] K. Yang, Q. Abbasi, K. Qaraqe, A. Alomainy, and Y. Hao, "Bodycentric nano-networks, EM channel characterisation in water at the terahertz band," in Asian Pacific Microwave Conference (APMC), Japan, November 2–5 2014, pp. 1–5.
[31] A. Sani, A. Alomainy, and Y. Hao, "Numerical characterization and link budget evaluation of wireless implants considering different digital human phantoms," IEEE Transactions on Microwave Theory and Techniques, vol. 57, pp. 2605–2613, October 2009.
[32] K. Sayrafian-Pour, W.-B. Yang, J. Hagedorn, *et al.*, "Channel models for medical implant communication," International Journal of Wireless Information Networks, vol. 17, pp. 105–112, December 2010.
[33] H. Bahrami, B. Gosselin, and L. A. Rusch, "Realistic modeling of the biological channel for the design of implantable wireless UWB communication systems," in Engineering in Medicine and Biology Society (EMBC) Annual International Conference, IEEE, 2012.

[34] A. Johansson, "Wireless communication with medical implants: Antennas and propagation," Ph.D. dissertation, Lund University, 2004.

[35] J. Kim and Y. Rahmat-Samii, "Implanted antennas inside a human body: Simulations, designs, and characterizations," IEEE Transactions on Microwave Theory and Techniques, vol. 52, no. 8, pp. 1934–1943, 2004.

[36] J. Gemio, J. Parron, and J. Soler, "Human body effects on implantable antennas for ISM bands applications: Models comparison and propagation losses study," Progress in Electromagnetics Research, vol. 110, pp. 437–452, October 2010.

[37] F. Merli, B. Fuchs, J. R. Mosig, and A. K. Skrivervik, "The effect of insulating layers on the performance of implanted antennas," IEEE Transactions on Antennas and Propagation, vol. 59, no. 1, pp. 21–31, 2011.

[38] K. Y. Yazdandoost and R. Kohno, "Wireless communications for body implanted medical device," in Asia-Pacific Microwave Conference, 2007.

[39] J. R. Reitz, J. M. Frederick, and R. W. Christy, Foundations of Electromagnetic Theory (4th ed.). Addison-Wesley, Reading. ISBN 0-201-52624-7, 1993.

[40] S. H. Lee, J. Lee, Y. J. Yoon, *et al.*, "A wideband spiral antenna for ingestible capsule endoscope systems: Experimental results in a human phantom and a pig," IEEE Transactions on Biomedical Engineering, vol. 58, pp. 1734–1741, June 2011.

[41] T. Karacolak, A. Hood, and E. Topsakal, "Design of a dual-band implantable antenna and development of skin mimicking gels for continuous glucose monitoring," IEEE Transactions on Microwave Theory and Techniques, vol. 56, pp. 1001–1008, April 2008.

[42] A. Laskovski and M. Yuce, "A MICS telemetry implant powered by a 27 MHz ISM inductive link," in Engineering in Medicine and Biology Society, EMBC, 2011 Annual International Conference of the IEEE, 2011.

[43] A. Kiourti and K. Nikita, "Miniature scalp-implantable antennas for telemetry in the MICS and ISM bands: Design, safety considerations and link budget analysis," IEEE Transactions on Antennas and Propagation, vol. 60, no. 8, pp. 3568–3575, August 2012.

[44] M. L. Scarpello, D. Kurup, H. Rogier, *et al.*, "Design of an implantable slot dipole conformal flexible antenna for biomedical applications," IEEE Transactions on Antennas and Propagation, vol. 59, no. 10, pp. 3556–3564, 2011.

[45] A. F. Demir, Q. H. Abbasi, Z. E. Ankarali, E. Serpedin and H. Arslan, "Numerical characterization of *in vivo* wireless communication channels," RF and Wireless Technologies for Biomedical and Healthcare Applications (IMWS-Bio), IEEE MTT-S International Microwave Workshop Series, London, pp. 1–3, 2014.

[46] Z. N. Chen, G. C. Liu, and T. S. See, "Transmission of RF signals between MICS loop antennas in free space and implanted in the human head," IEEE Transactions on Antennas and Propagation, vol. 57.6, pp. 1850–1854, 2009.

[47] L. C. Chirwa, P. Hammond, S. Roy, and D. R. S. Cumming, "Electromagnetic radiation from ingested sources in the human intestine between 150 MHz and 1.2 GHz," IEEE Transactions on Biomedical Engineering, vol. 50, pp. 484–492, April 2003.

[48] S. K. S. Gupta, S. Lalwani, Y. Prakash, E. Elsharawy, and L. Schwiebert, "Towards a propagation model for wireless biomedical applications," in IEEE International Conference on Communications (ICC), 2003.

[49] K. Y. Yazdandoost, K. Sayrafian-Pour, *et al.*, "Channel model for body area network (BAN)," IEEE P802, vol. 15, 2009.

[50] S. Stoa, C. S. Raul, and I. Balasingham, "An ultra wideband communication channel model for the human abdominal region," in GLOBECOM Workshops (GC Workshops), IEEE, 2010.

[51] "ANSYS HFSS®." [Online]. Available: http://www.ansys.com/Products/Electronics/ANSYS+HFSS®. [Accessed: 12-Nov-2015].

[52] A. Rahman and Y. Hao, "A novel tapered slot CPW-fed antenna for ultra-wideband applications and its on/off-body performance," in International Workshop on Antenna Technology: Small and Smart Antennas Metamaterials and Applications, 2007, IWAT '07., vol., no., pp. 503–506, 21–23 March 2007.

[53] Y. Liu and R. D. Gitlin, "A phenomenological path loss model of the *in vivo* wireless channel," in IEEE 16th Wireless and Microwave Technology Conference (WAMICON), April 2015.

[54] C. He, Y. Liu, G. E. Arrobo, T. P. Ketterl, and R. D. Gitlin, "*In Vivo* wireless communications and networking," in Information Theory and Applications Workshop (ITA), 2015, San Diego, CA, February 2015.

[55] A. F. Demir, Z. E. Ankarali, Q. H. Abbasi, Y. Liu, K. Qaraqe, E. Serpedin, H. Arslan, and R. D. Gitlin, "In Vivo Communications: Steps Toward the Next Generation of Implantable Devices," IEEE Vehicular Technology Magazine, vol. 11, no. 2, pp. 32–42, Jun. 2016.